\documentclass{l4dc2025}

\title[Short Title]{Online Voltage Regulation of Distribution Systems with Disturbance-Action Controllers}

\usepackage{times}
\usepackage{algorithm, algorithmic}
\usepackage{mathtools}
\usepackage{dsfont}
\usepackage{xcolor}
\usepackage{caption}

\coltauthor{\Name{Peng Zhang} \Email{pz233@uw.edu}\and
 \Name{Baosen Zhang} \Email{zhangbao@uw.edu}\\
 \addr Electrical and Computer Engineering, University of Washington, Seattle, WA 98195}

\begin{document}

\maketitle

\begin{abstract}%

Inverter-based distributed energy resources facilitate the advanced voltage control algorithms in the online setting with the flexibility in both active and reactive power injections. A key challenge is to continuously track the time-varying global optima with the robustness against dynamics inaccuracy and communication delay. In this paper, we introduce the disturbance-action controller by novelly formulating the voltage drop from loads as the system disturbance. The controller alternatively generates the control input and updates the parameters based on the interactions with grids. Under the linearized power flow model, we provide stability conditions of the control policy and the performance degradation to model inaccuracy. The simulation results on the radial distribution networks show the effectiveness of proposed controller under fluctuating loads and significant improvement on the robustness to these challenges. Furthermore, the ability of incorporating history information and generalization to various loads are demonstrated through extensive experiments on the parameter sensitivity.
\end{abstract}


\section{Introduction}
Modern distribution systems are integrated with high volumes of inverter-based distributed energy resources (DERs) like solar. These DERs can cause large voltage swings in the system. Conventional devices such as tap changing transformers and capacitor banks are mechanical devices and may not be able to correct the fast voltage variations introduced by the DERs~\cite{zhang2014optimal}. Therefore, using the power generated by the DERs themselves to control the voltage has been an active area of research.

Many voltage control algorithms only use reactive power, but because of the high $r/x$ ratio of the distribution systems, voltage is also sensitive to changes in active power~\cite{kersting2018distribution}. However, jointly optimizing both active and reactive power typically involves solving the optimal power flow (OPF) problem, which can be solved offline or online. 

Offline algorithms have been well-studied in the literature, for example, see \cite{molzahn2017survey} and the references therein. Those methods are based on fully solving an optimization problem with satisfaction of capacity and operation constraints. However, because of real-time uncertainties in the DERs, it is often difficult to find good offline solutions. Consequently, online algorithms have been proposed to explicitly exploit the physical grids as a solver to continuously obtain power setpoints. A gradient projection algorithm through implicit function theorem is adopted to pursue AC-OPF solutions in radial systems with the sufficient condition on convergence to global optima \cite{gan2016online}. For time varying loads, tracking errors has been studied in  \cite{tang2019time, simonetto2017prediction}. However, these methods are very sensitive to model uncertainties and communication delays. Since the distribution systems are not well monitored and wide area communications are not yet available~\cite{liao2015distribution,zhang2013local,feng2023stability}, model estimation errors and communication delays are sometimes unavoidable. 


The problem of model inaccuracy has certainly been a topic of focus in the control community. For example, robust control techniques such as $\mathcal{H}_\infty$ control or set-based control have been developed~\cite{zhou1998essentials,blanchini1999set}. However, these approaches are pessimistic and provide over conservative actions. Recently, a disturbance-based controller is used for non-stochastic case in the online manner \cite{agarwal2019online}. Specifically, such parameterization is obtained by a system-level approach to controller synthesis based on the closed-loop system response \cite{wang2019system}. It could bypass the non-convexity of state feedback controller and drive the policy to the global optima by gradient descent \cite{fazel2018global}.

\textbf{Contributions} 
Our main contribution is to design a feedback policy-based algorithm for online voltage regulation when the load information is dynamically changing in the distribution systems. At each time, the remote inverters implement the active and reactive power injections according to the control law. Meanwhile, the controller updates its parameter towards the global optima of designed loss function through the interactions with grids. We take our idea from the closed-loop system response based disturbance-action controller \cite{wang2019system}, by creatively treating time-varying loads as disturbances and explicitly exploiting the correlation characteristics. We also implement saturation on the control over running the algorithm, since all inverters have capacity limits. Furthermore, theoretical guarantee on the stability of controller is provided, mainly in terms of the magnitude of disturbance and solar energy. Simulation results show the significant improvement on the mitigation of over-voltage and robustness to the above challenges.

Our method is based on policy iteration rather than value iteration in those Lagrangian-based works. Our framework enables not only the learning process but also the extraction of intermediate policy and closed-form input over the interactions with physical grids. Furthermore, compared to the optimization techniques, more history information could be incorporated into the controller, especially when loads are high-correlated. In addition, faced with both the uncertain dynamics of distribution systems and latency, our algorithm could achieve over 50\% better robustness in contrast to directly solving the linearized AC-OPF problem in the real-time implementation. We note that  \cite{magnusson2020distributed} also demonstrates the robustness through simulation results by using primal-dual methods, although we are able to theoretically analyze the stability conditions and performance changes under the model inaccuracy.

\section{Models and Problem Formulation}
\subsection{Model of Distribution Systems}
Let us consider a radial distribution system, represented as a connected network $G = (\mathcal{N}', \mathcal{E})$, where $\mathcal{N}'=\{0,1,2, \ldots, n\}$ is the set of nodes and $\mathcal{E} \subset \mathcal{N}' \times \mathcal{N}'$ is the set of lines among them. The voltage phasor at bus $k$ is $v_k = |v_k| e^{j \angle v_k} \in \mathbb{C}$ and the injected current phasor is $i_k = |i_k| e^{j \angle i_k} \in \mathbb{C}$. We take bus 0 as the slack bus with fixed voltage $v_0$ (e.g. a feeder bus~\cite{kersting2018distribution}). 

The voltages and currents are linearly related as $i' = Y' v'$ via the nodal admittance matrix $Y' = \left[\begin{array}{cc}Y_{0} & Y_{\mathcal{N}} \\ Y_{\mathcal{N}}^T & Y\end{array}\right] \in \mathbb{C}^{(n+1)\times (n+1)}$, where $i' =\left[\begin{array}{c}
i_0 \\ i \end{array}\right] \in \mathbb{C}^{n+1}$ and $v' =\left[\begin{array}{c}
v_0 \\ v \end{array}\right] \in \mathbb{C}^{n+1}$. We are typically interested in the non-slack buses, i.e., the set of buses $\mathcal{N} = \{1, \dots, n \}$. By the Laplacian structure of $Y'$ and taking Schur's complement, we have~\cite{Low2024} 
\begin{align}
v_0=|v_0| e^{j \angle v_0}, \; 
i_0 = Y_{0} v_0 + Y_{\mathcal{N}}v, \; 
v = v_0 \mathds{1} + Z i,
\end{align}
where  $Z := Y^{-1}= R + jX$ for some real matrices $R$ and $X$. 

The complex power injection at the buses is defined as 
\begin{equation} \label{eqn:s}
    s=\operatorname{diag}\left(\bar{i}\right) v, 
\end{equation}
where $\bar{i}$ is the element-wise complex conjugate of $i$. Quite a bit of attention have been paid to linearize \eqref{eqn:s}, with the conclusion that under mild conditions, the voltage magnitudes can be written as 
\begin{align}
    |v|=\mathds{1}|v_0|+\frac{1}{|v_0|} \operatorname{Re}[Z \bar{s}].
\end{align}
We work with this standard linearized model and the interested reader can consult~\cite{bolognani2015existence,zhu2015fast} and the references therein. For notational simplicity, we drop the magnitude and write 
\begin{align}
    v_{t+1} = v_0 + \frac{1}{v_0}(Rp_t + Xq_t),
\end{align} 
with the understanding that lower case $v$ always represents the voltage magnitude.

We shift the system to look at the voltage deviation at time $t$ by defining $x_t$ as $x_t = v_t - v_0$.  Net active and reactive power injections are decomposed as non-controllable and controllable loads, represented by $p_t^l, q_t^l \in \mathbb{R}^n$  and $p_t^s, q_t^s \in \mathbb{R}^n$, respectively. We think of the controllable part as the input $u_t = [p_t^s; q_t^s] \in \mathbb{R}^{2n}$ and the non-controllable part as the disturbance $w_t = -\frac{1}{v_o}(Rp_t^l + X q_t^l) \in \mathbb{R}^{n}$. Together, we have a linear system of the form:
\begin{align}
    x_{t+1} = Bu_t + w_t,
    \label{system}
\end{align}
whose dynamics are $B = [R, X] \in \mathbb{R}^{n \times 2n}$. We note that the disturbance $w_t$ is not white, rather, the noises between different time steps are quite correlated~\cite{diagne2013review,chen2018model}. Physically, solar irradiation and loads do not change instantaneously, and as time steps become shorter because of advances in hardware (e.g., inverters), the disturbances generally change slower. In fact, our controller design for $u_t$ leverages this fact. 

The system in \eqref{system} is somewhat unusual, since $x_t$ does not appear on the right-hand side. The internal dynamics of the voltages on the distribution grids are quite fast (within a second), and longer time-scale variations are the result of the disturbance $w_t$. Several voltage control algorithms are based on the state feedback, where $u_t$ is a function of $x_t$~\cite{zhang2013local,feng2023stability} and the dynamics of $x_t$ would be in a more standard LTI equation. The controller in this paper is of a different form, which will be described in the next section. 


\subsection{Problem Formulation}
The online voltage control problem is to continuously drive the voltage deviation of each bus within the assigned safety interval, typically as $[\bar{x}, \underline{x}] \in \mathbb{R}^n$. In this paper, both inverter-based active and reactive power can be adjusted within the capacity limits, constrained by $p^s \in [\underline{p}^s, \bar{p}^s]$ and $q^s \in [\underline{q}^s, \bar{q}^s]$, i.e., $\underline{u} = [\underline{p}^s; \underline{q}^s]$ and $\bar{u} = [\bar{p}^s; \bar{q}^s]$. We are interested in the problem below:
\begin{subequations}
\begin{align}
    & \underset{u_t \in \mathbb{R}^{2n}}{\text{minimize}}
    &  &C_t(x_{t+1}, u_t) \\
    & \text{subject to} &  &x_{t+1} = B u_t + w_t \\
    &  & &u_t = f_t(x_t, \dots, x_1; u_{t-1}, \dots, u_0)\\
    & & &  \underline{u}_{t} \leq u_t \leq  \bar{u}_{t} \\
    & & & \underline{x}_{t+1} \leq x_{t+1} \leq  \bar{x}_{t+1}
\end{align}
\label{online control}
\end{subequations}
\hspace{-3pt}where $C_t(x_{t+1}, u_t) = C_t^x(x_{t+1}) + C_t^u(u_{t})$ is decomposed into two parts. $C_t^x(x_{t+1})$ is the system-level loss, such as voltage deviation, active power loss, \textit{etc}, while $C_t^u(u_t)$ is the control cost to reduce the solar energy curtailment (for more physical interpretations, see~\cite{zhu2015fast,feng2023stability}). We assume the above optimization problem is feasible, i.e., at each time $t$, existing power injection $u_t$ to satisfy the voltage limits. 


Explicitly solving \eqref{online control} might be difficult since, even with a linearized model, the problem is quite large and is sensitive to the inaccuracy and latency of the model. Therefore, as with most voltage control approaches, our goal is to design a feedback control law, $u_t = f_t(\cdot)$. 


\section{Algorithm and Main Results}
\subsection{Controller Design}
Here we introduce the design of our proposed disturbance-action controller on the real-time voltage control. In the closed-loop setting, the controller alternatively generates the control input and updates its parameter every time receiving the latest measurements of nodal voltage and implemented power injections from the power grids.


The concrete steps of the feedback control law are detailed in Algorithm 1. 
The algorithm starts from the initialization of disturbance collection that contains the perturbation information over the last $H$ steps, and controller parameter set whose elements determine the specific contribution of disturbance at corresponding time $t$. Then, with the appropriate choice of parameters, real-time voltage control is running by iteratively executing STEPS 2-4. Concretely, in STEP 2, the remote-installed inverters receive the maximum available active power $\bar{p}_t$ based on local weather condition and then compute the active and reactive power injections according to Equation (\ref{DAC}) through the information from aggregator. Next, in STEP 3, each inverter observes the new voltage deviation $x_{t+1}$ via the interaction with grids and send them to the aggregated controller together with implemented action $u_t$. Then, in STEP 4, the controller updates the disturbance collection with given measurements and estimated dynamics $\hat{B}$, and conducts one-step gradient descent on the parameter set.

The main intuition here is to explicitly exploit the auto-regressive load profiles via the special parameterization of feedback controller. To be precise, the disturbance-based control policy was introduced by \cite{wang2019system} as an alternative to the classic state/output feedback controller. Based on the closed-loop system response, it bypasses the nonconvexity of state feedback controller to the overall loss function and directly constructs the linear relation between system perturbation and control input. The sufficiency of such a class of policies to approximate any linear stabilizing policy was proved in [Lemma 5.2, \cite{agarwal2019online}].
In our setting, we leverage the fact that the noise is correlated in time to achieve disturbance rejection.  Furthermore, it guarantees that the steepest step is taken towards the global optima by gradient descent operator at each time $t$. 

In the process of updating disturbance sequence and controller parameters, we make the following two assumptions:

\textit{Assumption 1.} The model dynamics are bounded by $\| B\|\leq \kappa_B$, and inaccuracy from learned model is under $\epsilon_B$, i.e., $\|B - \hat{B} \| \leq \epsilon_B$.

\textit{Assumption 2.} Cost function $C_t$ is convex and Lipschitz continuous with constant $L$ for $(x, u)$, i.e, $\|\nabla_{x} C_t(x_t, u_t) \|$, $\|\nabla_{u} C_t(x_t, u_t) \| \leq L D $ as long as $\|x_t \|$, $\|u_t\| \leq D, \; \forall t$.

As stated in \textit{Assumption 1}, we do not assume that we know the exact model, instead we assume that the error is not too large.  
So it is useful to think of a surrogate system defined as
\begin{align}
  \hat{x}_{t+1} = \hat{B}u_t + \hat{w}_t,  
\end{align}
which differs the state variable $x_{t+1}$ in model and disturbance. The loss function for system-level and controller-level costs corresponds to the one in Equation (\ref{online control}) with satisfaction of \textit{Assumption 2}.

To ensure that the problem is well-posed, we need the following two assumptions: 

\textit{Assumption 3.} The disturbance sequence is set as auto-regressive series and bounded by $\|w_t \| \leq W$.

\textit{Assumption 4.} The original uncontrolled input $\Tilde{u}_t$ is a random variable independent of others and bounded by $\| \Tilde{u}_t \| \leq \Tilde{U}$.




\begin{algorithm}
\caption{Disturbance-Action Controller for Online Voltage Control}
\begin{algorithmic}
\STATE \textbf{STEP 1} (Initialization) 
\STATE \textbf{input:} Estimated dynamics $\hat{B}$, Tracing horizon $H$, Learning rate $\eta$, Parameters $W, \Tilde{U}, \epsilon_B, \gamma$
\STATE \textbf{initialize:} Disturbance collection $\{(\hat{w}_{t-1}, \dots, \hat{w}_{t-H}) | \hat{w}_t=0, t < 0\}$, Controller parameter $M^{[0]} = \{M_1^{[0]}, \dots, M_H^{[0]}\}, || M_i^{[0]}|| \leq \frac{2\Tilde{U}}{\epsilon_B \Tilde{U} + W} \gamma^i, (i = 1, \dots, H)$
\WHILE{ $t\geq 0$}
    \STATE \textbf{STEP 2} (Local Implementation)
    \STATE Receive the latest available solar energy $\bar{p}_t$ and obtain natural input $\Tilde{u}_t = \left[\begin{array}{c} \bar{p}_t \\ 0 \end{array}\right]$
    \STATE Compute and saturate the actual active and reactive power injection
    \begin{align}
    u_t = \left[\Tilde{u}_t + \sum_{i=1}^{H}{M_i^{[t]}\hat{w}_{t-i}}\right]_{\underline{u}_t}^{\bar{u}_t}
        \label{DAC}
    \end{align}
    \STATE \textbf{STEP 3} (Data Collection) Observe the new voltage deviation $x_{t+1} = Bu_t + w_t$
    \STATE \textbf{STEP 4} (Controller Update)  Update disturbance collection $\hat{w}_t = x_{t+1} - \hat{B} u_t$
    \STATE Update controller parameter 
         $M^{[t+1]} = M^{[t]} - \eta \nabla_{M}C_t(\hat{x}_{t+1}(M), u_t(M))$
    \STATE $t = t+1$
\ENDWHILE
\label{DAC_alg}
\end{algorithmic}
\end{algorithm}

\subsection{Main Results}
Here we state and prove the main theoretical results on the performance of the controller and its robustness to model inaccuracies. For simplicity, we assume the one-step tracing horizon here, i.e., $H=1$, while the conclusion can be easily generalized to the longer horizons and numerical experiments are conducted for different length of tracing windows in Section 4.


\textit{Theorem 5.} Under \textit{Assumptions 1-4}, it is sufficient to achieve stability and avoid fluctuation over both state and input variables when no estimation errors are imposed on dynamics, i.e., $\epsilon_B = 0$, by setting initialization of controller as $\|M^{[0]} \| \leq \frac{2 \Tilde{U}}{W}$ and learning rate as
\begin{align}
    \eta \leq \frac{2 \Tilde{U}}{LDd (1 + \kappa_B) W^2},
\end{align}
where denotes by the dimensionality of the problem $d = \text{max}\{\text{dim}(x_t), \text{dim}(u_t)\}$.


The theorem illustrates that the stability is guaranteed provided that the step size is small enough. To interpret the upper bound, it is proportional to the solar generation ($\Tilde{U}$) and inverse proportional to the square of voltage drop from loads ($W^2$). This relation adapts to the scenario of high PV penetration, where a less conservative step size is required. Moreover, stability is maintained even under the communication delay as long as the consecutiveness of observations holds at the controller.


\begin{proof}
Note that because of the operational limits on both active and reactive power, $p_t^s$ 
and $q_t^s$, the measured voltage deviation $x_t$ and control input $u_t$ are bounded, according to dynamics (\ref{system}). Thus, the stability here is established on the controller parameters $M^{[t]}$, which is tantamount to bounding the variation of controlled input between consecutive time steps,
\begin{align}
    \|M^{[t]}\hat{w}_{t-1} - M^{[t+1]}\hat{w}_{t} \| \leq 2 \| \Tilde{u}_t\|.
    \label{stability}
\end{align}
Using assumptions aforementioned, we have that
\begin{align*}
    \|M^{[t]}\hat{w}_{t-1} - M^{[t+1]}\hat{w}_{t} \| \leq \|M^{[t]} - M^{[t+1]} \| \|\hat{w}_t \|
    & \leq \eta \|\nabla_{M}C_t(\hat{x}_{t+1}(M), u_t(M)) \| \|\hat{w}_t \| \\
    & \leq \eta LDd (W + \kappa_B W) W,
\end{align*}
where the first inequality is by Cauchy-Schwartz and last inequality is from \textit{Lemma 6} (given below). Then, criterion on step size is obtained by assigning the upper bound in Equation (\ref{stability}). 

Regarding the initialization, since $\hat{w}_t = 0$ if $t<0$, the above expression is reduced to 
$\| M^{[1]} \hat{w}_0\|$ $\leq \|M^{[1]} \| W \leq 2 \Tilde{U}$ or $\Longrightarrow \| M^{[1]}\| \leq \frac{2 \Tilde{U}}{W}$.
Together with the following triangle inequality on $\|M^{[0]} \|= \|M^{[0]} - M^{[1]} + M^{[1]} \| \leq \|M^{[0]} - M^{[1]} \| + \|M^{[1]}\|$, it is sufficient for the initial parameterization to choose the same value. 
\end{proof} 

 \textit{Lemma 6.} (A bound on the norm of policy gradient) Under \textit{Assumptions 1-4}, the bound on the norm of gradient in estimation-free case, i.e, $\epsilon_B = 0$ and $w_t = \hat{w}_t$, is obtained as follows, 
\begin{align}
  \|\nabla_{M}C_t(\hat{x}_{t+1}(M), u_t(M)) \|  \leq LDWd (1 + \kappa_B).
\end{align}

\begin{proof}
For error-free scenario, no computation inaccuracy is incurred and surrogate variable $\hat{x}_{t+1}$ is reduced to true state variable $x_{t+1}$. Next, regarding matrix variables, the $\mathcal{L}_2$ norm of the gradient corresponds to the Frobenius norm of gradient matrix $\nabla_M C_t$. Thus, we start by considering the bound on single element. Follow the initialization in \textit{Theorem 5}, we get that the initial state and control are bounded, where $\|x_{t+1}\|$ $\|u_t\| \leq D$, with $D= \text{max}\{2 \Tilde{U}, \kappa_B \Tilde{U} + W\}$. This comes from $\| x_{t+1} \| \leq \|B\| \| \Tilde{u}_t\| + \|w_t \| \leq \kappa_B \Tilde{U} + W$ and $\| u_t\| \leq 2 \| \Tilde{u}_t \| \leq 2 \Tilde{U}$.

Then, according to \textit{Assumption 2} and chain rule, we have that
\begin{align*}
    |\nabla_{M_{p, q}}C_t(\hat{x}_{t+1}(M), u_t(M)) |  \leq L D (\|\frac{\partial \hat{x}_{t+1}(M)}{\partial M_{p, q} } + \frac{\partial u_t(M)}{\partial M_{p, q} }\|) 
    &\leq LD (\|\frac{\partial \hat{x}_{t+1}(M)}{\partial M_{p, q} }\| + \|\frac{\partial u_t(M)}{\partial M_{p, q} }\| )\\
    & \leq LD (|w_{q,t}| + \|B\| |w_{q,t}|),
\end{align*}
where $w_{q,t}$ denotes the \textit{q-th} element on $w_t$. The last inequality is obtained through the linear relationship between disturbances and state and input, which is the main distinction of disturbance-action controller to the classic state/output feedback control.
Then, by adding all elements together, we get that 
\begin{align*}
  \|\nabla_{M}C_t(\hat{x}_{t+1}(M), u_t(M)) \| \leq LDd (\|w_t\| + \|B\| \|w_t\|) \leq LDWd (1 + \kappa_B).
\end{align*}
\end{proof}

Next, we extend the stability condition to the case when model is inaccurate. This is necessarily stricter, with smaller controller parameters and step size. The proof follows the similar steps in \textit{Theorem 5} with only replacement of the new bounds on perturbations. Specifically, since the estimation error directly influences the calculation of disturbance, we can obtain the new bound as $\|\hat{w}_t \|\leq \epsilon_B \Tilde{U} + W$ by deriving that
\begin{align}
    \hat{w_t}&= x_{t+1} - \hat{B} u_t= Bu_t + w_t - \hat{B}u_t \nonumber = (B - \hat{B})u_t + w_t.
\end{align}

\textit{Theorem 7.} Under \textit{Assumptions 1-4}, it is sufficient to achieve stability on the state and input variables with model estimation error as $\|B - \hat{B} \| \leq \epsilon_B$, by choosing initialization of controller as $\|M^{[0]} \| \leq \frac{2 \Tilde{U}}{\epsilon_B \Tilde{U} + W}$ and  learning rate as
\begin{align}
    \eta \leq \frac{2 \Tilde{U}}{LDd (1 + \kappa_B) (\epsilon_B \Tilde{U} + W)^2}.
\end{align}

Based on the stability guarantee, we can quantify the performance variation of proposed controller faced with model inaccuracy. This is achieved by the bounded controller parameters guaranteed by the above theorems, i.e, $\|M^{[i]} \| \leq \bar{M}$. The discrepancy in state, input, and calculated disturbance at each time $t$ are defined by $\Delta u_t$, $\Delta x_t$, and $\Delta 
\hat{w}_t$, respectively.

\textit{Theorem 8.} Suppose that the disturbance-action controller is implemented with the stability condition on $\eta$ and $\| M^{[0]}\|$, and the estimation error is bounded by $\epsilon_B \leq \frac{W}{\Tilde{U}}$. Then, it holds true that $\|\Delta u_t \| \leq \Bar{Y}_t$ and $\|\Delta x_{t+1} \| \leq \Bar{X}_{t+1}$, where $\Bar{Y}_t$ and $\Bar{X}_{t+1}$ are defined as follows,
\begin{align}
    \Bar{Y}_t & \triangleq \begin{cases} 
(\Bar{M}(\kappa_B + \epsilon_B))^{t-1}\| \Delta u_1\| & \text{if } \Bar{M}(\kappa_B + \epsilon_B) \leq 1, \\
\text{min}\{(\Bar{M}(\kappa_B + \epsilon_B))^{t-1}\| \Delta u_1\| , \Tilde{U}\} & \text{if } \Bar{M}(\kappa_B + \epsilon_B) > 1,
\end{cases}\\ 
 \Bar{X}_{t+1} & \triangleq \kappa_B \Bar{Y}_t,
\end{align}
with the base case calculated as $\|\Delta \hat{w}_0 \| \leq \Tilde{U} \epsilon_B$ and $\| \Delta u_1\| \leq \Bar{M} \Tilde{U} \epsilon_B$. 

\begin{proof}
To bound the performance degradation, let us start from the base case. With $\hat{w}_t = 0$ ($t<0$), we get that $\Delta u_0 = 0$, and $\Delta \hat{w}_0= (B - \hat{B}) u_0 $ by following the dynamics aforementioned. Thus, we have that $ \|\Delta \hat{w}_0 \| \leq \Tilde{U} \epsilon_B. $
Next, with the proper initialization in \textit{Theorem 7} and the requirement of model inaccuracy as $\epsilon_B \leq \frac{W}{\Tilde{U}}$, it is to be verified that $\epsilon_B \|M^{[1]} \| \leq 1$. Thus, we get that
\begin{align*}
    \| \Delta u_1\| \leq \text{min}\{\|M^{[1]}\| \|\Delta \hat{w}_0 \|, \Tilde{U}\} \leq \Bar{M} \Tilde{U} \epsilon_B.
\end{align*}
The subsequent upper bounds can be validated by continuously expanding the equations of state evolution whereby the induction. Note that the condition on $\Bar{M}(\kappa_B + \epsilon_B) \leq 1$ denotes the contraction on the performance variations, while activating the saturation limits at any instant will result in $\Bar{Y}_t = \Tilde{U}$ in the all following steps.
\end{proof}

\begin{figure}[ht!]
    \centering
    \begin{minipage}{0.55\textwidth}
        \captionsetup{aboveskip=10pt, belowskip=-10pt}
        \centering
        \includegraphics[width=\linewidth]{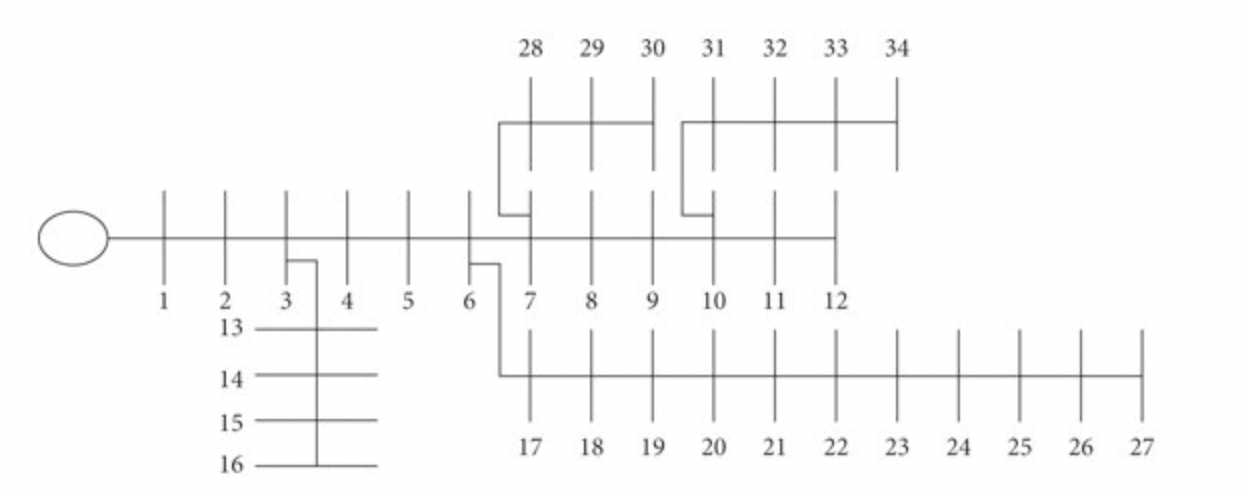}
        \caption{One-line diagram of 34-node redial distribution network}
        \label{feeder}
    \end{minipage}
    \hfill
    \begin{minipage}{0.4\textwidth}
        \captionsetup{aboveskip=0pt, belowskip=-10pt}
        \centering
        \includegraphics[width=\linewidth]{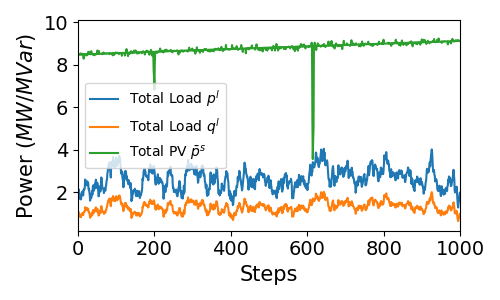}
        \caption{Total load and PV generation profiles. The downward spikes result from the sharp decrease in PV output (e.g., from cloud cover). }
        \label{data}
    \end{minipage}
\end{figure}

\section{Simulation Results}
In this section we present case studies of the proposed controller on the 34-bus radial distribution network. For simplicity, we assume that the system operation is balanced three-phase (or equivalently, single-phase). The detailed line parameters are provided in \cite{salama1993simplified} and the topology is shown in Figure \ref{feeder}. The simulations are divided into three parts. First, we demonstrate the effectiveness of Algorithm (\ref{DAC_alg}) to keep bus voltages within the safety limits under scenarios with high PV penetration and time-varying loads. Second, the robustness against both model inaccuracy and latency is shown. Specifically, we compare our method to explicitly solving the optimization problem where the model is inaccurate. Third, we study the sensitivity analysis on the tracing horizon $H$ and the correlation coefficient $\alpha$. 

\subsection{Environment Setup}
We assume that three different sizes of PV-inverters are randomly installed at a subset of buses \footnote{Specifically at buses $\{2, 3, 6, 10, 12, 13, 14, 16, 17, 19, 24, 31, 32, 34\}$}.
To mimic the sudden shading from cloud cover and other fast events, the available active power can experience large drops. Regarding the exogenous load profiles $p^l_t$ and $q^l_t$, we generate them from an auto-regressive series (see Figure \ref{data}), i.e, $p^l_{t+1} (q^l_{t+1})= \sqrt{1 - \alpha} p^l_t (q^l_{t}) + \sqrt{\alpha} \eta_t$, where $\eta_t \sim \mathcal{N}\left(0, \sigma^2 I_n\right)$ denotes the uncontrollable load variations and $\alpha$ is the correlation coefficient (set as 0.1 in the simulations).

The nominal voltage magnitude at each bus is 11 kV and the safety interval takes the standard value of $\pm 5\%$:$[\underline{x}_t, \bar{x}_t] =[-0.55\text{kV}, 0.55\text{kV}]$~\cite{zhang2014optimal}. At the inverter level, we assume that it can curtail the output power below the maximum available value but cannot absorb the active power, resulting in the bound to be $[\underline{p}^s_t, \bar{p}^s_t] = [0, \bar{p}^s_t]$. As for reactive power, we assume that an inverter can produce or absorb it up to some limit, which we set to be $40\%$ of the total power, i.e. $[\underline{q}^s_t, \bar{q}^s_t] = [-0.4\bar{p}^s_t, 0.4 \bar{p}^s_t]$. For buses without inverters, both the upper and lower bounds are set to zero. The objective function $C_t$ is designed to minimize the real power curtailment and reactive power production, expressed as $C_t= c_p\|p^s_t - \bar{p}^s_t\|^2 + c_q \|q^s_t \|^2$, where $c_p = 3$ and $c_q = 1$.

To simulate model uncertainty, we consider both parametric and topological errors. Each line impedance is multiplied by a random scaling factor sampled from $[0.8, 1.2]$ uniformly. Then, the positions of four buses are permuted randomly. Consequently, the estimated dynamics $\bar{B}$ used here have a relatively error of about $25\%$. Finally, note that all experiments are completed on a MacBook Pro with an Apple M1 chip and 8 GB RAM and through Auto-diff tool (TensorFlow) for gradient calculation.

\subsection{Performance Evaluation}
\begin{figure}[ht!]
    \centering
    \begin{minipage}{0.3\textwidth}
    \captionsetup{aboveskip=-1pt, belowskip=-5pt}
        \centering
        \includegraphics[width=\linewidth]{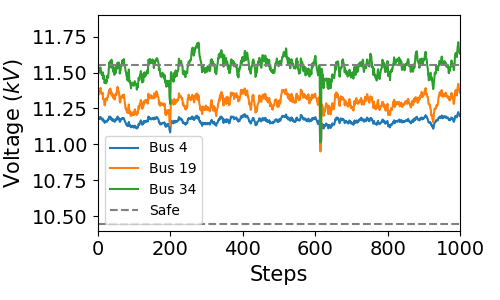}
        \caption*{(a) No control}
    \end{minipage}
    \hfill
    \begin{minipage}{0.3\textwidth}
    \captionsetup{aboveskip=-1pt, belowskip=-5pt}
        \centering
        \includegraphics[width=\linewidth]{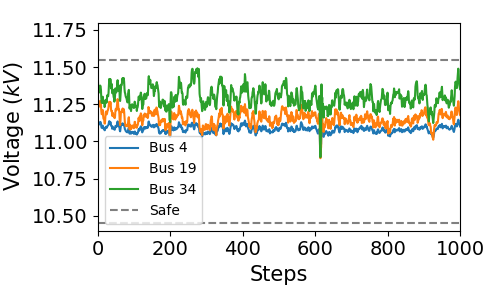}
        \caption*{(b) Disturbance-action controller}
    \end{minipage}
    \hfill
    \begin{minipage}{0.3\textwidth}
    \captionsetup{aboveskip=-1pt, belowskip=-5pt}
        \centering
        \includegraphics[width=\linewidth]{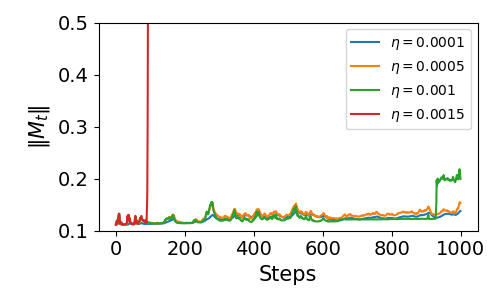}
        \caption*{(c) Controller parameter}
    \end{minipage}
    \caption{Voltage profiles with different controllers. The nominal voltage is 11 kV with a tolerance of $\pm 5\%$. When there is no control (Fig. (a)), the voltage at some buses can exceed their limits. Our proposed controller (Fig. (b)) keeps all the voltages within their required bounds. Fig. (c) shows the importance of the $\eta$ parameter, as a large value would case the system to be unstable.}
    \label{main1}
\end{figure}
Figure \ref{main1} shows the ability of the proposed algorithm to reliably maintain the nodal voltage within the safety intervals under the dynamics estimation error. The voltage profiles at buses 4, 19, 34 are shown, where the maximum voltage deviation is achieved at the end of the network, that is, bus 34. If no control is done, then over-voltages would occur.  With our controller, all voltages remain within the safety intervals even after sudden and large PV changes around step 200 and 600. Notice that only one past step information is incorporated here ($H=1$) and to drive the voltage approaching the nominal value, the regulation term ($\| x_{t+1}\|^2$) is added with weights as $c_x = 0.5$. In addition, the choice of learning rate and controller parameter follows the guidance of \textit{Theorem 7}, with $\eta = 5 \times 10^{-4}$ and  $M^{[0]} = \begin{bmatrix}
0.05I_n \\
0.1I_n
\end{bmatrix}$.  The importance of the learning rate is shown in Figure \ref{main1} (c), where the inappropriate choice of step size ($\eta=0.0015$) will lead to instabilities. In contrast, the controller can continuously update the parameter to track the influence of time-varying loads and solar profiles under the stability guarantee. Finally, the running time per trajectory is 7.7 milliseconds, which is fast enough to be applied in practice and leave sufficient time for the convergence of inverter-level dynamics.

\begin{figure}[ht!]
    \centering
    \begin{minipage}{0.45\textwidth}
        \captionsetup{aboveskip=-1pt, belowskip=-5pt}
        \centering
        \includegraphics[width=\linewidth]{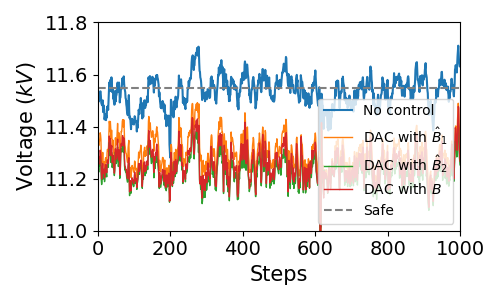}
        \caption*{(a) Disturbance-action controller}
    \end{minipage}
    \hfill
    \begin{minipage}{0.45\textwidth}
        \captionsetup{aboveskip=-1pt, belowskip=-5pt}
        \centering
        \includegraphics[width=\linewidth]{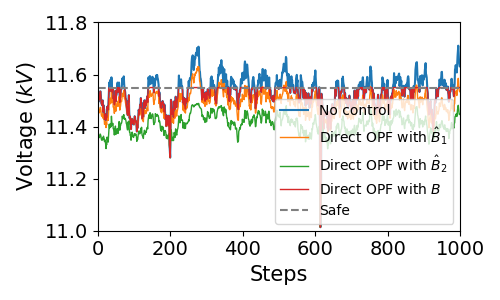}
        \caption*{(b) Direct optimization method}
    \end{minipage}
    \caption{Voltage profiles of the robustness against model inaccuracy with different controllers. The proposed controller (Fig. (a)) achieves the low sensitivity to two randomly chosen inaccurate estimations, while explicitly solving the optimization model (Fig. (b)) results in the significant performance variations.}
    \label{inaccuracy}
\end{figure}
\textit{Robustness to model inaccuracy.} Figure \ref{inaccuracy} illustrates that the proposed algorithm is less sensitive to inaccurate estimations than directly solving the optimization model. Suppose that the true model is $B$ and two estimations of it are $\hat{B}_1$ and $\hat{B}_2$. The proposed controller is able to satisfy the safety constraints no matter which model is used and the voltage profiles are roughly the same even when the wrong model is used. However, explicit optimization can lead to large errors if the model is wrong. Given that the distribution systems are not particularly well known and well-monitored~\cite{yu2017patopa}, robustness to model uncertainty is beneficial to practical implementations. 

\begin{figure}[ht!]
    \centering
    \begin{minipage}{0.45\textwidth}
        \captionsetup{aboveskip=-1pt, belowskip=-5pt}
        \centering
        \includegraphics[width=\linewidth]{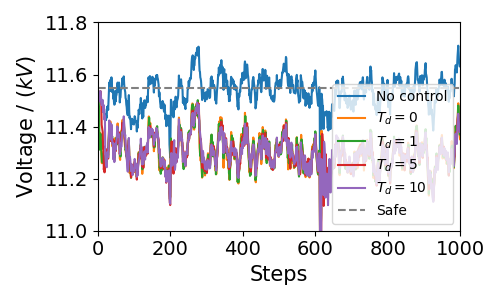}
        \caption*{(a) Overall voltage information}
    \end{minipage}
    \hfill
    \begin{minipage}{0.45\textwidth}
        \captionsetup{aboveskip=-1pt, belowskip=-5pt}
        \centering
        \includegraphics[width=\linewidth]{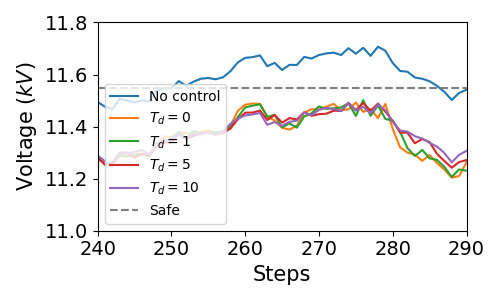}
        \caption*{(b) Local information over steps: 240-290}
    \end{minipage}
    \caption{Voltage profiles of the robustness against communication delay with the disturbance-action controller. Maintaining the voltage within safety limits, a tolerance to the large two-way latency ($T_d = 10$) is exhibited in Fig. (a) and (b), regarding the overall state evolution and detailed local information, respectively.}
    \label{latency}
\end{figure}


\textit{Robustness to latency.} Figure \ref{latency} depicts the robustness of our controller to the communication delay. Specifically, the centralized controller receives the feedback from grids $T_d$ steps before and transmits the updated parameter to the remote inverters with another $T_d$-step delay. In Figure \ref{latency} (b), a closer look is taken for the detailed information over certain steps. It is noticed that the voltage profiles entangle each other and have no significant differences from the latency-free scenario, under all four distinct delay levels ($T_d = 1, 5, 10$).


\subsection{Parameter Sensitivity}
We demonstrate the ability of disturbance-action controller to incorporate more history information through the sensitivity analysis on the tracing horizon $H$, since primal-dual gradient methods only leverage one step previous observations. For numerical results, three evaluation metrics are adopted: average voltage deviation ($\frac{1}{N} \sum^{T}_{t=1}\|x_t\|^2$), total control cost ($\sum^{T}_{t=1}C_t$) and nodal voltage fluctuation ($\frac{\mu_i}{\sqrt{\frac{1}{T} \sum_{t=1}^{T} (x_{i,t} - \mu_i)^2}}$, $\mu_i = \frac{1}{T} \sum_{t=1}^{T} x_{i,t}$).

\begin{figure}[ht!]
    \centering
    \begin{minipage}{0.24\textwidth}
        \captionsetup{aboveskip=-1pt, belowskip=-5pt}
        \centering
        \includegraphics[width=\linewidth]{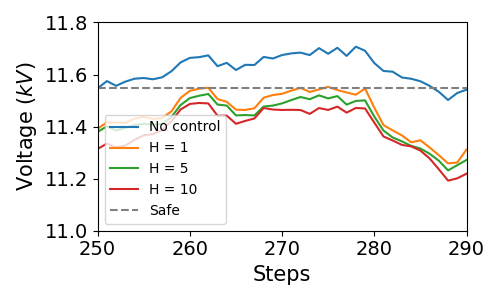}
        \caption*{(a) Voltage profile}
    \end{minipage}
    \begin{minipage}{0.24\textwidth}
        \captionsetup{aboveskip=-1pt, belowskip=-5pt}
        \centering
        \includegraphics[width=\linewidth]{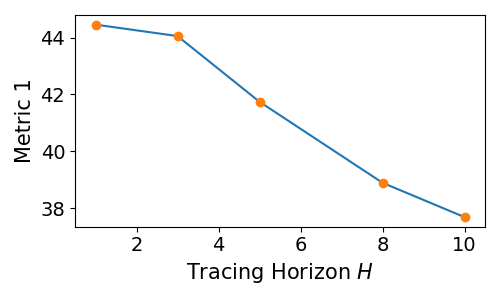}
        \caption*{(b) Voltage deviation}
    \end{minipage}
    \begin{minipage}{0.24\textwidth}
        \captionsetup{aboveskip=-1pt, belowskip=-5pt}
        \centering
        \includegraphics[width=\linewidth]{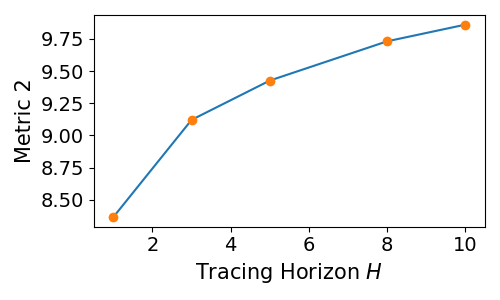}
        \caption*{(c) Control cost}
    \end{minipage}
    \begin{minipage}{0.24\textwidth}
        \captionsetup{aboveskip=-1pt, belowskip=-5pt}
        \centering
        \includegraphics[width=\linewidth]{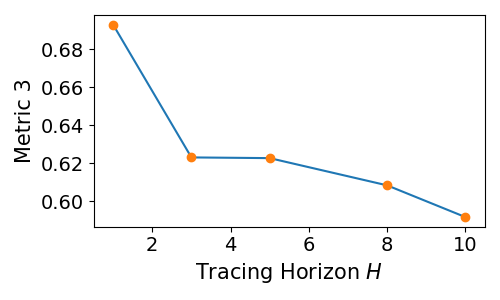}
        \caption*{(d) Voltage fluctuation}
    \end{minipage}
    \caption{Sensitivity analysis on the tracing horizon $H$. Fig. (a) illustrates the voltage variations at bus 34 under different $H$, where incorporating history observations can enhance the controller performance. Three metrics in terms of voltage deviation, control cost and voltage fluctuation are utilized to demonstrate the conclusion quantitatively.}
    \label{sensitivity}
\end{figure}

The performance variations to increasing tracing horizons are shown in Figure \ref{sensitivity}, with a  learning rate of $\eta = 10^{-4}$. A longer horizon leads to a better voltage performance, but at higher costs. This is a trade-off that can be decided by operators based on practical needs.  


\begin{figure}[ht!]
    \centering
    \begin{minipage}{0.3\textwidth}
        \captionsetup{aboveskip=-1pt, belowskip=-5pt}
        \centering
        \includegraphics[width=\linewidth]{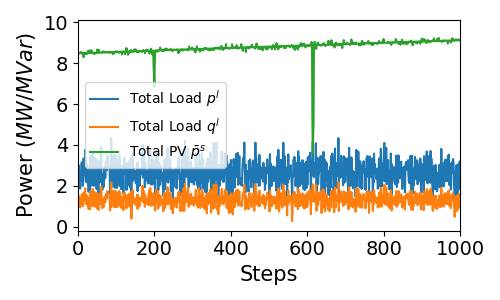}
        \caption*{(a) Total loads with $\alpha=1.0$}
    \end{minipage}
    \hfill
    \begin{minipage}{0.3\textwidth}
        \captionsetup{aboveskip=-1pt, belowskip=-5pt}
        \centering
        \includegraphics[width=\linewidth]{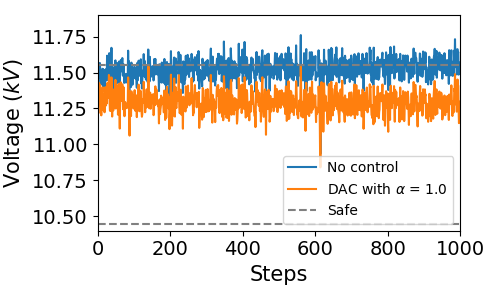}
        \caption*{(b) Voltage profile at bus 34}
    \end{minipage}
    \hfill
    \begin{minipage}{0.3\textwidth}
    \captionsetup{aboveskip=-1pt, belowskip=-5pt}
        \centering
        \includegraphics[width=\linewidth]{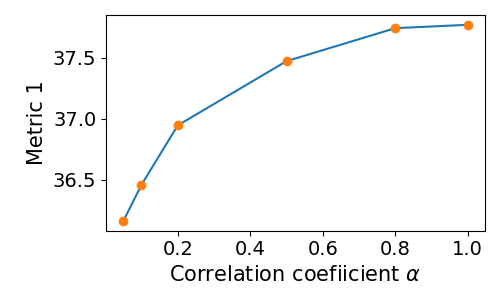}
        \caption*{(c) Voltage deviation}
    \end{minipage}
    \caption{Sensitivity analysis on the correlation coefficient $\alpha$. When load data is generated stochastically, i.e., $\alpha=1.0$ in Fig. (a), the nodal voltage still can be almost regulated by our proposed controller (Fig. (b)). The performance degradation is quantified by the numerical metric on voltage deviation, shown in Fig. (c).}
    \label{correlation}
\end{figure}

Finally, we illustrate the generalization of our controller under various loads, based on the sensitivity analysis of the correlation coefficient $\alpha$. The performance changes to decreasing correlation are exhibited in Figure \ref{correlation}. Although the voltage deviation amounts as the correlation reduces, our controller is able to regulate the voltage safely when $\alpha=1.0$, with only $4.4\%$ degradation to the nominal case. Our future work is to incorporate realistic data to further evaluate the controller.

\section{Conclusion}
This paper considers the online voltage control of distribution systems under the time-varying loads and solar profiles. The disturbance-action controller is introduced to maintain the nodal voltage within safety intervals through the interactions with girds. Compared to directly solving a specific optimization problem, our proposed framework enjoys the better robustness against the model estimation error and communication delay. In addition, the stability of controller and corresponding performance variation are theoretically provided by the proper choice of parameter update. The ability of incorporating history data and generalizing to various loads are demonstrated through the simulations of parameter sensitivity. In future work, we will work on decomposing the calculated disturbances by parts to further account for the linearization error, obtainable load predictions and uncontrollable accidents.




\end{document}